# Ultrasensitive Self-powered large area planar GaN UV-photodetector using reduced graphene oxide electrodes

Nisha Prakash[1,2], Manjri Singh[1,2], Gaurav Kumar[1], Arun Barvat[1,2], Kritika Anand[1,2], Prabir Pal[1,2,*], Surinder P. Singh[1], Suraj P. Khanna[1,*]

[1]CSIR-National Physical Laboratory, Dr. K. S. Krishnan Road, New Delhi 110012, India

[2]Academy of Scientific and Innovative Research, CSIR-National Physical Laboratory (Campus), Dr. K. S. Krishnan Road, New Delhi 110012, India

A simplistic design of a self-powered UV-photodetector device based on hybrid r-GO/GaN is demonstrated. Under zero bias, the fabricated hybrid photodetector shows a photosensivity of ~ 85% while ohmic contact GaN photodetector with identical device structure exhibits only ~5.3% photosensitivity at 350 nm illumination (18 µW/cm$^2$). The responsivity and detectivity of the hybrid device were found to be 1.54 mA/W and 1.45×10$^{10}$ Jones (cm Hz$^{½}$ W$^{−1}$), respectively at zero bias with fast response (60 ms), recovery time (267 ms) and excellent repeatability. Power density-dependent responsivity & detectivity revealed ultrasensitive behaviour under low light conditions. The source of observed self-powered effect in hybrid photodetector is attributed to the depletion region formed at the r-GO and GaN quasi-ohmic interface.

The tremendous progress in gallium nitride (GaN) based light emitting diodes (LEDs),[1] laser diodes[2], and other GaN based devices, namely, UV photodetector (PD) has attracted a great deal of interest from research community.[3] Being chemically inert and thermally stable, they are the most suitable for applications such as flame detection, secure space communication and ozone layer monitoring.[4,5] To match the requirements for such applications in remote and extreme environment, it is highly desirable for the UV devices to be ultrasensitive, with fast response and operate in self-powered mode. However, inherent high defect density associated with as-grown epitaxial GaN films limits its performance.[6]

The various schemes of metal-semiconductor (M-S) interfaces have been used to improve the UV-PDs device performance. The use of two different metal electrodes on n-GaN with modulating Schottky barrier height leads to a fast response speed but limited reverse saturation current density.[7] Earlier study has shown Schottky contact photodiode of Ni/GaN/Au with asymmetric interdigitated finger electrodes having a responsivity of 5 mA/W in self-powered mode at UV illumination.[8] Recently, Sun *et al.* have reported a high responsivity of 104 mA/W at zero bias voltage of interdigitated Schottky contact photodiode of Ni (80 nm) /GaN/Ti (20 nm) /Al (60 nm) device.[9] On the other hand, the use of highly transparent conductive electrodes (TCE) such as indium-tin-oxide (ITO), cadmium-tin-oxide (CTO) on n-GaN improves the reverse saturation current density while the forward bias current density is marginal.[10] It is also reported that a thicker interfacial layer forms at the interface of ITO/GaN during fabrication process which increases the contact resistance.[11] Moreover, the M-S contacts are implemented with interdigitated finger electrodes that require a large number of processing steps and turns out be complicated and expensive.

Reduced-graphene-oxide (r-GO) has emerged as an alternative to ITO that can provide low contact resistance along with large-scale applicability for next generation PD applications with relatively low-cost and simplified method.[12,13] In addition, numerous thermal and chemical reduction methods have been employed to provide optical and electrical tunability of r-GO, as reported in previous studies.[14] The synthesis of r-GO nanosheets is scalable and cost effective and has been proposed to revolutionize the nanoelectronics domain because of its unique properties such as decent charge mobility, high thermal conductivity, excellent electronic conductivity, high mechanical strength, and bandgap tunability.[15]

Few research groups have reported the study of the effect of integration of r-GO with GaN technology. Ryu *et al.* have shown improved light-current-voltage performance of GaN LED by utilising r-GO as a current spreading layer.[16] Han *et al.* discussed the role of r-GO as a heat spreading layer to enhance the thermal stability of high-power GaN LEDs.[17] Recently, Liday *et al.* have demonstrated low resistance contacts by using CNT/r-GO/CNT on p-type GaN.[18] Typically, a Schottky contact PD is used for self-powered applications. However, there is no report yet demonstrating the utility of r-GO as TCE forming non-ohmic (quasi) contact on GaN epitaxial film. In this letter, we conclusively demonstrate a simple fabrication process of r-GO/GaN hybrid structure where r-GO forms a quasi-ohmic contact instead of ideal ohmic contact on unintentionally-doped (UID) epitaxial GaN film. It is observed that the fabricated PD operates in self-powered mode with much improved photosensitivity compared with that of ohmic contact GaN PD operated with an applied bias voltage. The self-powered photodetector also exhibits fast response and recovery time with low dark current. The results suggest that r-GO integrated with GaN is a promising candidate for self-powered UV PD applications.

Firstly, UID-GaN film was grown on 2-inch *c*-plane sapphire substrate (0001) by plasma-assisted molecular beam epitaxy (PAMBE) technique. Prior to loading into vacuum, standard chemical processes were used to clean the sapphire substrate. The growth was carried out at a substrate temperature of 657 °C, RF plasma power of 425 W, and beam equilibrium pressures for gallium and nitrogen were $6.87\times10^{-7}$ Torr and $5.1\times10^{-5}$ Torr, respectively. *In-situ* reflection high energy electron diffraction (RHEED) technique was used to monitor the growth conditions to achieve smooth morphology of GaN film. The defect density in the as-grown

GaN film is ~$1.06\times10^9$ cm$^{-2}$ obtained from rocking curve full width at half maximum (FWHM) for (0002) plane of GaN. Secondly, the GO was synthesized using modified Hummers method[19] in conjunction with freeze drying to achieve monodispersed and highly exfoliated GO nanosheets. These nanosheets were then uniformly dispersed in de-ionised (DI) water (0.5 mg/mL) using ultrasonication. To fabricate r-GO/GaN lateral heterojunction the GaN (5 mm × 5 mm) sample was cleaned thoroughly by the standard degreasing solvents, rinsed in DI water and then blow dried. GO solution was then drop casted onto GaN film to form parallel electrodes with a gap of 800 μm. The fabricated GO/GaN heterostructure was allowed to dry in air and then annealed in ambient conditions at 300°C for 30 mins to form thermally reduced r-GO. The size of each electrode was estimated to be ~ 2.1 mm × 5.0 mm. For undertaking electrical measurements copper wires were connected to the two r-GO electrodes through conducting silver paint with contact area diameter of 1.0 mm. Using similar dimensions, an ohmic contact GaN PD (without r-GO) was fabricated with indium as a large area electrodes.

UV-Visible absorption spectra of GaN, r-GO and r-GO/GaN samples were measured using UV-Vis-NIR spectrophotometer (CARY5000). The current-voltage (I-V) characteristic curves were measured using a Keithley source meter (2635B) under dark and light illuminations. Photoconductivity measurements were performed in dc mode using continuous illumination of weak monochromatic light at different wavelengths. The light wavelengths used were varied from 250 to 450 nm by using a TMc 300 monochromator of 0.1 nm resolution integrated with Xenon / Quartz-Halogen lamps. The incident power density at each wavelength was measured by using a calibrated Si-DH photodetector. In order to study the optical switching behaviour of the fabricated device a mechanical shutter was employed to investigate the light ON-and-OFF behaviour.

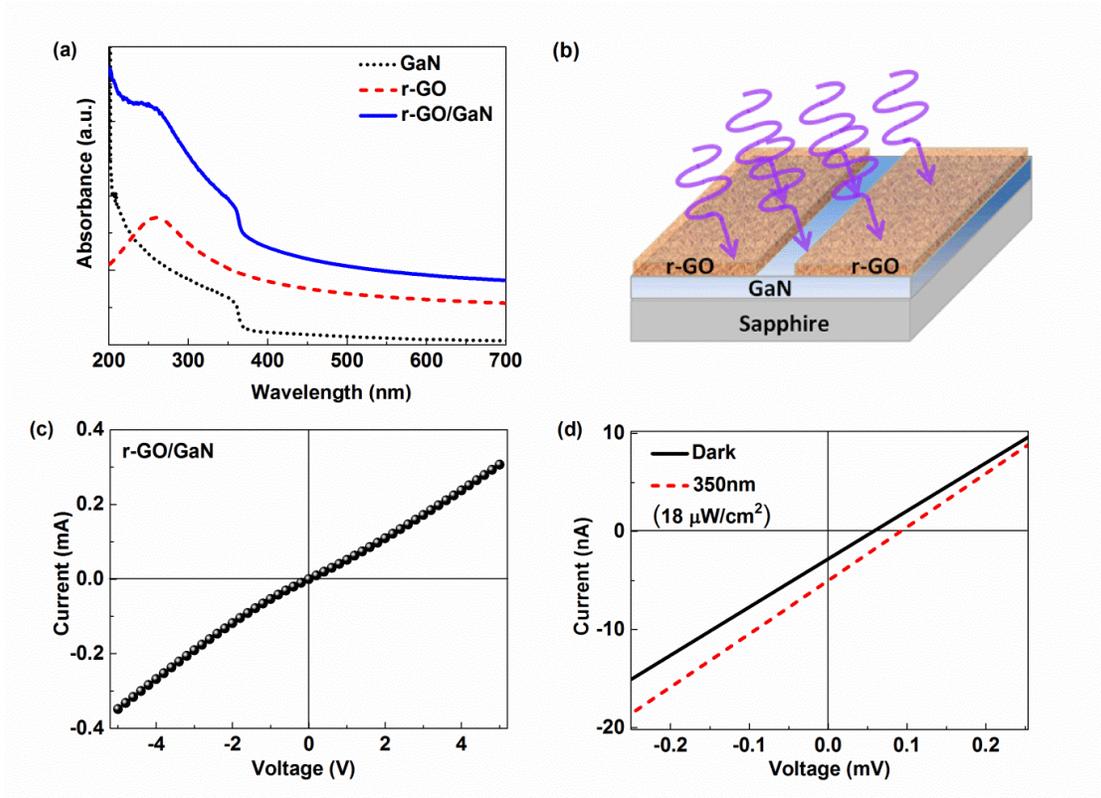

FIG.1. (a) UV-Visible absorption spectra of GaN, r-GO and r-GO/GaN heterostructure, (b) schematic illustration of r-GO/GaN PD with an effective active area of ~ 21 mm$^2$ out of total area of 25 mm$^2$, (c) current-voltage characteristic of the fabricated r-GO/GaN PD under dark condition, and (d) narrow range current-voltage characteristics of the r-GO/GaN PD under dark ($V_{oc}$ ~ 0.05 mV) and illuminated ($V_{oc}$ ~ 0.1 mV) conditions showing a photovoltaic behavior.

Figure 1 (a) shows the UV-visible absorption spectra of the GaN, r-GO and r-GO/GaN heterostructure. It can be seen that the absorption edge of the GaN is located at 365 nm and is associated with the direct band gap of GaN.[20] While bare r-GO shows the absorption peak at 270 nm related to π→π* transition in r-GO.[21] Interestingly, r-GO/GaN heterostructure exhibits a strong absorption region in the UV range from 270-365 nm indicating that this heterostructure could act as wide range UV-PD. Figure 1 (b) and 1 (c) shows the schematic diagram and I–V characteristics curve under dark conditions of the r-GO/GaN PD, respectively. The I-V curve exhibits that r-GO forms a quasi-ohmic contact with GaN. A separate I-V measurement revealed that silver forms a good ohmic contact with r-GO (not shown). This clearly depicts that the observed quasi-ohmic behaviour in the r-GO/GaN PD is arising from the r-GO/GaN interface. Figure 1(d) shows the narrow range I-V curves of r-GO/GaN PD under dark and illumination with a wavelength of 350 nm with power density of 18 μW/cm$^2$. The device exhibits small photovoltaic behavior with an open-circuit voltage ($V_{oc}$) ~ 0.1 mV which enables the device to operate without any external bias voltage.

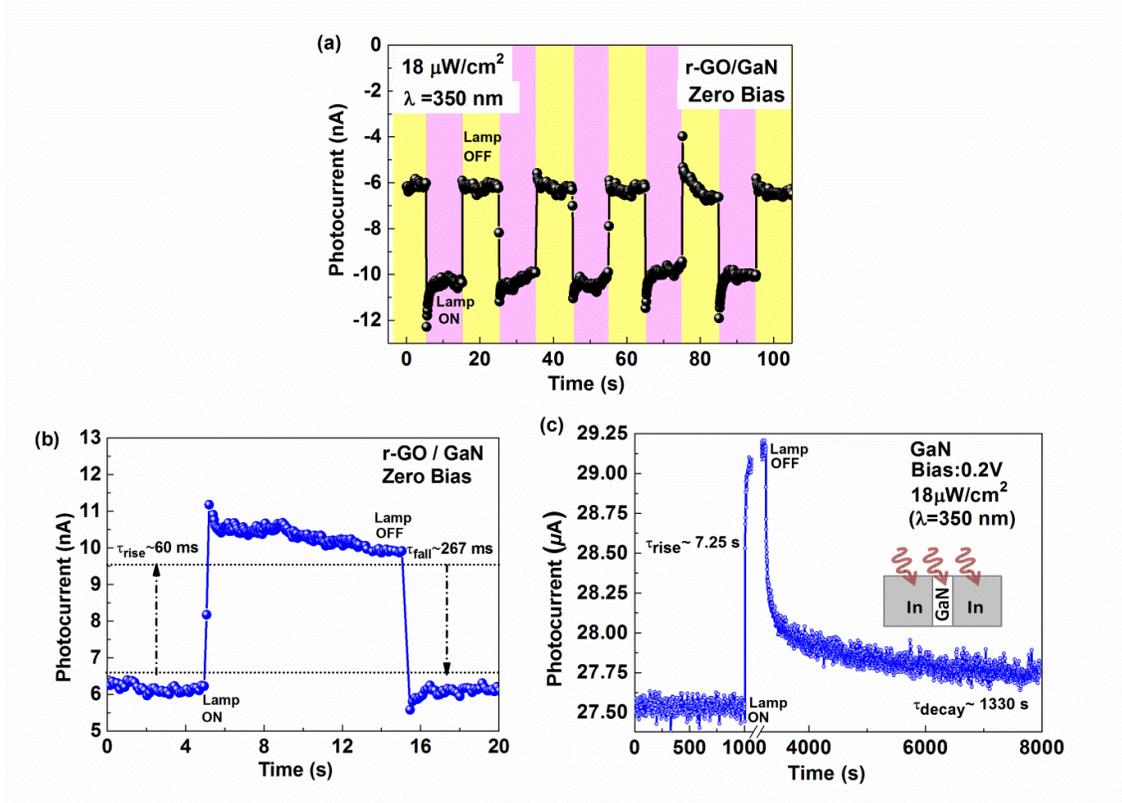

FIG.2. (a) Optical switching performance of the quasi-ohmic r-GO/GaN PD upon illumination of 350 nm wavelength. (b) Absolute value of single ON/OFF cycle showing $\tau_{rise}$ and $\tau_{fall}$ of quasi-ohmic r-GO/GaN PD, and (c) Photoresponse of ohmic-contact GaN PD with indium electrodes having an effective active area of ~4.0 mm² out of total area of 25 mm² showing $\tau_{rise}$ and $\tau_{decay}$.

Figure 2(a) shows the optical switching behaviour of r-GO/GaN PD under illumination of 350 nm wavelength with power density of 18 µW/cm². The device exhibits a sharp ON/OFF transition in photocurrent with good stability and reproducibility upon light pulse exposure at zero bias voltage. Figure 2(b) presents the magnified view of one of the absolute value of y-axis photocurrent switching cycle, revealing very fast photoresponse of the PD. The response and recovery speeds of the device are termed as rise time ($\tau_{rise}$) and fall time ($\tau_{fall}$) where, $\tau_{rise}$ is estimated from 10 to 90% of maximum photocurrent and $\tau_{fall}$ from 90 to 10% of maximum photocurrent.[22] The estimated $\tau_{rise}$ and $\tau_{fall}$ values of r-GO/GaN device are ~60 and 267 ms, respectively. The fast rise and fall in photocurrent in r-GO/GaN device has been attributed to the effect of r-GO as large area TCE where the transportation of photo-carriers dramatically increased and, thereby enhanced the speed of the device. The photocurrent of the device was calculated using,

$$I_{ph} = I_{Light} - I_{Dark} \tag{1}$$

where, $I_{Dark}$ is the dark current and $I_{Light}$ is the current under illumination. We observed a very low dark current of ~6.2 nA at zero bias voltage which is required to have high sensitivity. Whereas for GaN PD with indium electrodes, $\tau_{rise}$ and $\tau_{decay}$ are 7.25 and 1329 s, respectively as shown in Figure 2(c). It is evident from the figure that the GaN itself is not a very fast PD which may be attributed to the higher density of trapped electronic states due to induced defects in the material.[23,24] The quasi-ohmic r-GO/GaN device exhibits a photosensitivity ($I_{ph}/I_{Dark}$) of ~ 85% under self-powered mode with 350 nm illumination (18 mW/cm$^2$). In contrast, the ohmic contact GaN device with indium electrodes does not exhibit any photoresponse at zero bias voltage. The photosensitivity is found to be 5.3% in case of ohmic contact GaN photodetector with 350 nm illumination (18 mW/cm$^2$). Thus, the photosensitivity of r-GO/GaN self-powered PD is enhanced by ~16 times compared to that of ohmic contact GaN PD operated at 0.2 V bias.

To evaluate the detection performance of r-GO/GaN PD, the responsivity ($R_\lambda$) and detectivity ($D^*$) are further estimated. The responsivity is defined as the ratio of photocurrent to the power density of incident light on the effective active area of the device. Similarly, the detectivity is a figure of merit and describes the lower limit to which a PD can respond. The two parameters can be expressed as

$$R_\lambda = \frac{I_{ph}}{P_{in} A} \tag{2}$$

$$D^* = \frac{R_\lambda \sqrt{A}}{\sqrt{(2e I_{Dark})}} \tag{3}$$

where, $P_{in}$ is the power density of incident light, $A$ is the effective active area and $e$ is the electronic charge.

Figure 3(a) shows the responsivity as a function of incident light over the UV region operated at zero bias voltage. The peak responsivity is found to be ~ 1.67 mA/W illuminated with 340 nm wavelength with an incident power density of 17.2 μW/cm$^2$. The peak is found to be blue shifted as compared to the GaN absorption peak observed at 365 nm (Fig. 1(a)), attributed to the increased band gap of the r-GO/GaN heterostructure. The observed good responsivity of r-GO/GaN PD at zero bias voltage is useful for low power dissipation applications. Further, the r-GO/GaN PD was tested under varied incident power density at a fixed wavelength of 350 nm at zero bias, as shown in Figure 3(b). The photocurrent increases as we increase the incident power density and shows a sub-linear dependence on it. This phenomenon demonstrates that the amount of photogenerated charge carriers is proportional to the incident light flux. The dependence of generated photocurrent on incident power density can be fitted by a power law: $I_{ph} \alpha P_{in}^{\theta}$ where, $P_{in}$ is the power density of incident light, and $\theta$ is a parameter related to the trapping and recombination processes of the photogenerated carriers in PD.[25] The value of $\theta$ is found to be 0.64 for r-GO/GaN device, suggesting the involvement of complex processes of electron-hole generation, recombination and trapping at the effective active area of r-GO/GaN interface.[25] The figure also depicts the responsivity as a function of incident power density. The responsivity of the device increases rapidly from 4.08 to 9.23 mA/W when the incident power density changes from 3.86 to 1.05 µW/cm$^2$. However, an increase of power density from 3.86 µW/cm$^2$ onwards leads to a saturation in the responsivity. The saturation of this photoresponse may be attributed to filling of effective trap states leading to the reduction of photoconductive gain. For power density below 4.0 µW/cm$^2$, the observed sharp increased responsivity suggests an ultrasensitive behaviour at low light intensity. The increased photosensitivity of the device is attributed to the domination of long-lived trap states providing high photoconductive gain at low optical powers.[26,27] Figure 3(c) shows the responsivity and detectivity plotted together in the same graph. Under 350 nm illumination with power density of 1.05 µW/cm$^2$, we have estimated detectivity to be 8.18x10$^{10}$ Jones (cm Hz$^{1/2}$ W$^{-1}$).

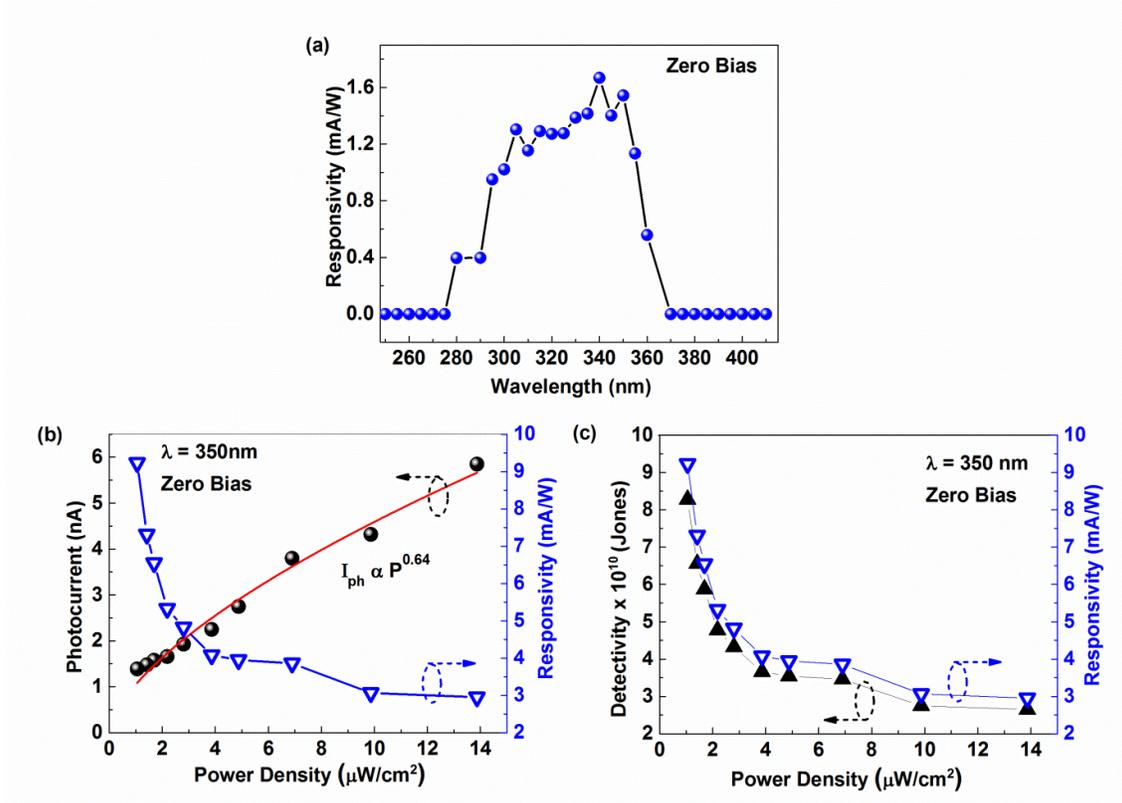

Fig. 3. (a) Responsivity of the r-GO/GaN PD over the UV region operated at zero bias voltage, (b) photocurrent and responsivity as a function of incident power density, and (c) responsivity and detectivity as a function of incident power density.

To understand the self-powered r-GO/GaN device and the mechanism of photocurrent generation and transportation, we propose an energy band diagram with the equivalent circuit, as displayed in Fig. 4 (a), (b). The difference between the work functions of r-GO (4.66 eV)[28] and GaN (4.2 eV)[29] leads to the formation of depletion region at the r-GO/GaN interface. This depletion region exists over the whole lateral interface between the two r-GO electrodes and the GaN film. Furthermore, the entire depletion region at the interface is directly illuminated through the two transparent large area r-GO electrodes. At these interfaces, two unlike built-in electric fields from the GaN to r-GO surface (GaN with a lower work function than r-GO) develops due to inhomogeneous nature of the drop casted electrodes as confirmed by the photovoltaic behaviour of quasi-ohmic contact of r-GO with GaN. This built-in electric field at a given light intensity dissociates the photogenerated excitons at the two interfaces. The difference between the two unlike built-in electric fields leads to a net electric field in the device which allows the prompt transportation of photo induced carriers. This geometry provides a large effective active area interacting with the light, thereby enhancing the overall photosensitivity of PD. Upon illumination with UV light the photogenerated carriers at the r-GO/GaN interface drift under the influence of net electric field even under zero external bias condition, as shown in the Fig. 4.

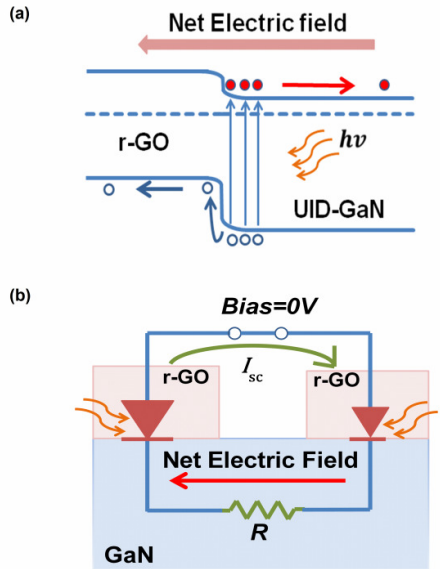

Fig. 4. (a) Energy band diagram depicting the generation and movement of photo-carriers at one of the r-GO/GaN interface, when the PD is operated at zero bias voltage. Under the influence of net electric field electrons move towards right while holes move towards left where they are collected separately by the respective r-GO electrodes, (b) equivalent circuit of r-GO/GaN self-powered PD.

In contrast to quasi-ohmic r-GO/GaN PD, the ohmic contact GaN PD has a different mechanism of photo carrier generation and transport of electron-hole pairs under UV light illumination. Here, the photo generated electrons and holes are spatially separated by the potential barrier created by the intrinsic defects in GaN. With the application of external bias voltage, the electrons and holes are drifted to the electrodes and gives rise to photocurrent. After turning off the UV illumination, the electrons and holes are trapped by the defect states and their wavefunctions become spatially localized with reduced overlapping, and thereby preventing the recombination with increased lifetime of the carriers. Thus, a long decay of 1330 s is observed in ohmic contact GaN PD, which is known as persistent photoconductivity. In view of realizing large area planer quasi-ohmic r-GO/GaN self-powered PD, the presence of conducting r-GO in the hybrid device facilitates the effective charge separation at interface and allows fast migration of photo carriers through r-GO.

In conclusion, we have demonstrated the integration of transparent r-GO electrodes with GaN by utilizing a simple drop casting technique with simplified device fabrication process, reduced cost as well as processing time. The r-GO/GaN contact displays a quasi-ohmic behavior. This fabricated self-powered hybrid device exhibits high photosensitivity (85%) and fast photoresponse ($\tau_{rise}$ ~ 60 ms) and recovery times ($\tau_{fall}$ ~ 267 ms) over the UV region. The present scheme could be utilized to improve the performance of UV-PDs and may be extended to applications in other self-powered opto-electronic devices or sensors.

The authors would like to thank Director, CSIR-NPL for encouragement. The work was funded by CSIR, India under OLP120132. S. P. K. and G. K. thanks DST Start Up Research Grant (Young Scientists), GAP150832, India for financial support. S. P. S. and M. S. acknowledges the CSIR network project NanoSHE (BSC 0112) for financial support. N. P. and A. B. thanks UGC, India for financial support towards their Ph. D.